# High Order Delta-Sigma Modulation with Positive Integer Coefficients

1st Martin J. W. Schubert
*Fakultät Elektro- und Informationstechnik*
*Ostbayerische Technische Hochschule (OTH) Regensburg*
Regensburg, Germany
martin.schubert@oth-regensburg.de

*Abstract*—This document proposes binomial integer parameters for the cascaded ΔΣ-modulator structure with distributed feedback and distributed feedforward input and multi-bit output. It is demonstrated that high orders can be achieved with these coefficients. Accuracy requirements concerning the coefficients are discussed.

## I. INTRODUCTION

Delta-Sigma (ΔΣ) modulators translate high resolution data to lower resolution data with higher sampling rate such, that higher resolution can be regained by lowpass filtering. Integer number coefficients in high-order modulators are useful for analog-to-digital (ADCs) and digital-to-digital converters (DDCs), which are the basis for ΔΣ digital-to-analog converters (DACs). Nowadays, multistage cascaded, noise-shaping modulators (MASH) are popular due to their simplified stability [1], [2], [3]. High order, defined as order >2, is particularly critical for single bit output [4], [5].

In ΔΣ-ADCs, that are typically realized with switched capacitor (SC) technology [6], coefficients are realized as ratios of capacitors, which are most accurately fabricated as integer multiples of a unit capacitor, so that all non-linear edge-effects of the unit capacitor are multiplied, as illustrated in Fig. 1.

ΔΣ-DACs are based on an inner ΔΣ DDC as illustrated in Fig. 2(d). It reduces the bit-width of a digital data stream from $m$ to $n < m$ after the input data rate was increased from $f_1$ to $f_2 > f_1$. Also in the DDC integer coefficients are advantageous, as they can be multiplied faster and more energy efficient than floating-point numbers, particularly in low-power processors as used e.g. in wireless sensor nodes.

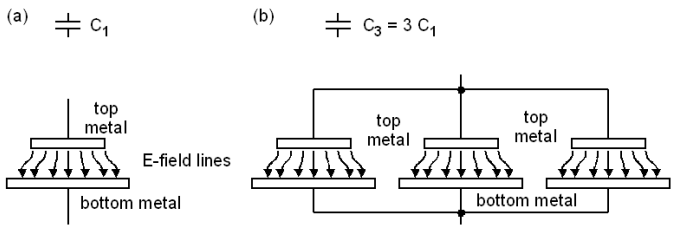

Fig. 1. Physical realization of capacitors (a) $C_1$ and (b) $C_2 = 3\ C_1$ as integer multiple of a unit capacitor to cancle out its nonlinear edge effects.

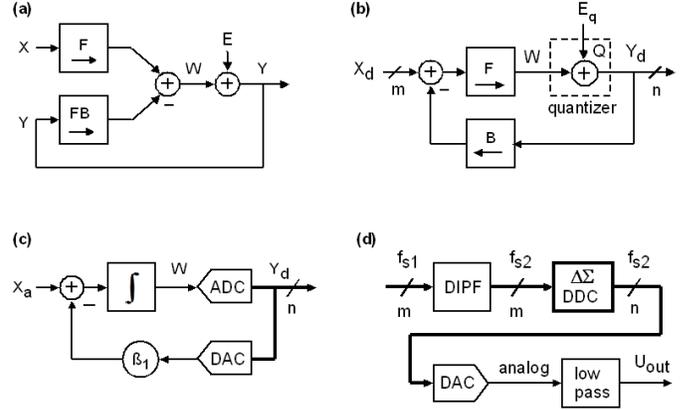

Fig. 2. (a) control loop principle with forward and loop networks $F$ and $FB$, respectively, (b) linear networks may share parts of the structure, here $F$, (c) analog-to-digital and (d) digital-to-analog converting ΔΣ modulators.

ΔΣ data converters are constructed as linear feedback loops, as shown in Fig. 2(a), where the quantity $W$ is a function of the two signals $X$ and $Y$

$$W(z) = F \cdot X + FB \cdot Y \qquad (1)$$

A further condition required to close the loop is

$$Y(z) = E + W \qquad (2)$$

Assuming $FB = F \cdot B$ allows for the definition of the feedback network $B = FB / F$ which is a rather mathematical construct in the case of distributed feedback as illustrated in Fig. 3.

For the closed loop, signal transfer function (*STF*) and noise transfer function (*NTF*) are defined combining (1) and (2).

$$STF = \left.\frac{Y}{X}\right|_{E_q=0} = \frac{F}{1+FB} \xrightarrow{|FB|\to\infty} \frac{1}{B}, \qquad (3)$$

$$NTF = \left.\frac{Y}{X}\right|_{E=0} = \frac{1}{1+FB} \xrightarrow{|FB|\to\infty} 0. \qquad (4)$$

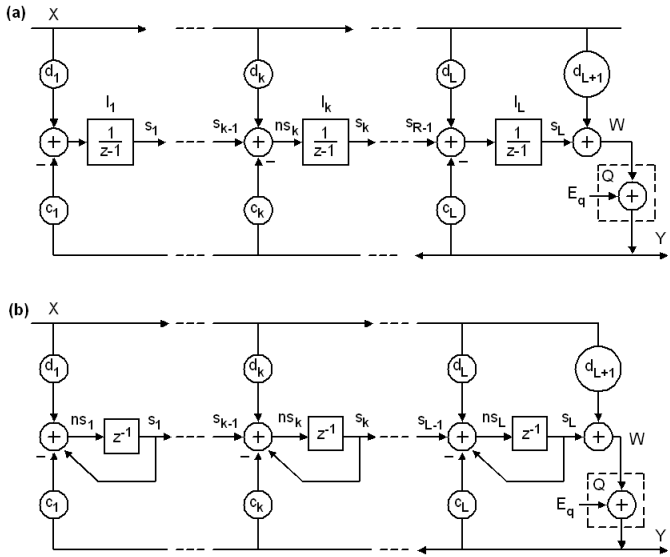

Fig. 3. *CIFB* structure: $L$ cascaded, time-discrete, delaying integrators with distributet feedback and distributed feedforward input, (a) symbolic integrators, (b) representation for implementation in simulation and digital hardware.

Feedforward network $F$ and loop-gain $FB$ are defined as

$$F = [W / X]_{Y=0} \quad (5)$$

$$FB = [W / Y]_{X=0}. \quad (6)$$

Application of (5), (6) to the $L$ cascaded, time-discrete, delaying integrators illustrated in Fig. 3 delivers

$$F = \left.\frac{W}{X}\right|_{Y=0} = \sum_{k=1}^{L+1} \frac{d_k}{(z-1)^{L+1-k}}, \quad (7)$$

$$FB = \left.\frac{W}{Y}\right|_{X=0} = \sum_{k=1}^{L} \frac{c_k}{(z-1)^{L+1-k}}, \quad (8)$$

Using the denominator polynomial $D(z)$

$$D(z) = (z-1)^L + \sum_{k=1}^{L} c_k (z-1)^{k-1}, \quad (9)$$

the noise transfer function is given in (4.34) of [2] as

$$NTF = \frac{1}{1+FB} = \frac{(z-1)^L}{D(z)}. \quad (10)$$

The Signal transfer function is achieved from $STF = F \cdot NTF$.

$$STF = \frac{d_1 + d_2(z-1) + ... + d_{L+1}(z-1)^L}{D(z)} \quad (11)$$

## II. THE PROPOSED METHOD

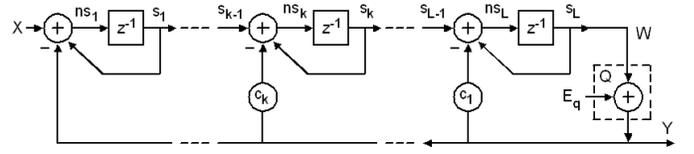

Fig. 4. *CIFB* structure using $L$ cascaded integrators with distributet feedback.

Fig. 4 illustrates the circuit of Fig. 3 for the particular case $d_1 = 1$ and $d_k = 0$ for $k > 1$. Loop gain $FB$ and consequently the $NTF$ are are according to (6), (8) and (10) no function of the parameters $d_k$. The signal transfer function becomes in this case

$$STF = \frac{1}{D(z)}. \quad (12)$$

If the $c_k$ are computed as binomial parameters

$$c_k = \binom{L}{k}, \quad (13)$$

Then the denominator polynomial $D(z)$ according to (9) is

$$D(z) = z^L \quad (14)$$

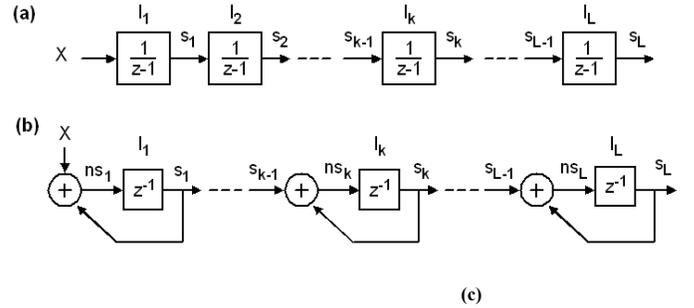

(a) Cascade of $L$ symbolic time-discrete integrators.

(b) Cascade of integrators, representation for implementation in simulation and hardware development.

(c) State values that develop as impulse responses on state vector $s = [s_1 ... s_6]$ form binomial coefficients.

| $n$ | $s_1$ | $s_2$ | $s_3$ | $s_4$ | $s_5$ | $s_6$ |
|---|---|---|---|---|---|---|
| 1 | 1 | 0 | 0 | 0 | 0 | 0 |
| 2 | 1 | 1 | 0 | 0 | 0 | 0 |
| 3 | 1 | 2 | 1 | 0 | 0 | 0 |
| 4 | 1 | 3 | 3 | 1 | 0 | 0 |
| 5 | 1 | 4 | 6 | 4 | 1 | 0 |
| 6 | 1 | 5 | 10 | 10 | 5 | 1 |
| 7 | 1 | 6 | 15 | 20 | 15 | 5 |
| 8 | 1 | 7 | 21 | 35 | 35 | 21 |
| 9 | 1 | 8 | 28 | 56 | 70 | 56 |
| 10 | 1 | 9 | 36 | 84 | 126 | 126 |

Fig. 5. Cascaded integrators, (a) symbolic and (b) detailed, generate binomial coefficients as shown in (c).

and according to (11)

$$STF(z) = z^{-L}. \quad (15)$$

The signal transfer function reduces to a simple delay of $L$ clock periods, while the noise transfer function has $L$ zeros at $z_{n,k} = 1$ for $k = 1…L$. Although more favorable zeros are presented, in e.g. in [2], this choice of parameters for $c_k$ allows for modulators with high orders. An intuitive explanation is provided by the fact that after clearing all memories, binomial coefficients emerge as impulse response at the outputs of the cascaded, time-discrete delaying integrators, as illustrated in Fig. 5(c). *Matlab* function *f_Pacsal* in Fig. 6 models Fig. 5 [7].

```
function pascal = f_Pascal(L);
% function gener. binomial params: Pascal's pyramide
% L: input: order of (a+b)^L
% pascal: output vector of length L+1
% Editor: Martin J.W. Schubert, Date: 24. May 2025
x=zeros(1,L+1);x(1)=1; % impulse for impulse response
s=zeros(1,L+1); ns=s; % state vecor
for n=1:L+1;
  % evaluate next-state vector
  ns(1)=s(1)+x(n);
  ns(2:L+1)=s(1:L)+s(2:L+1);
  % latch next-state vector into state-memory
  s=ns;
end;
pascal=s; % write to output vector
```

Fig. 6.  Listing of the *Matlab* function for binomial parameters of $(a+b)^L$.

```
function [y,e]=f_ds_cifb(x,L,d,c,dq,llim,ulim);
% function to evaluate Delta-Sigma mod. for CIFB type
% N: number of samples on y, e, x
% y(1:N): output: quantized output signal
% e(1:N): output: quantization error
% x(1:N): input: samples input values at time tn
% L: input: modulator order
% d(1:L+1): input: feedforward input coupling coefs
% c(1:L):   input: feedback network coefs
% dq: input: quantization step size delta-q
% llim: input: lower limit of y(), default -inf
% ulim: input: upper limit of y(), default +inf
% Martin Schubert, 22.May.2025
y=zeros(1,length(x));
s=zeros(1,L); ns=s; % state and next-state vectors
yn=0; eq=0; % initialize first samples
for n=1:length(x); % loop over samples
  % evaluate next-state vector
  ns(1)=s(1)+d(1)*x(n)-c(1)*yn;
  for k=2:L;
    ns(k)=s(k-1)+s(k)+d(k)*x(n)-c(k)*yn;
  end;
  s=ns; % apply active clock edge: state <- nextstate
  w=s(L)+d(L+1)*x(n);%compute signal before quantizer
  if dq>0;
    yn=dq*round(w/dq); % quantize
  else
    yn=w;
  end;
  if n<L; eq=0; else eq=yn-x(n-L+1); end;
  yn=max(min(yn,ulim),llim); % limit output
  y(n)=yn; e(n)=eq; % write data to output vectors
end;
```

Fig. 7.  Listing of the *Matlab* function modeling the *CIFB* structure of Fig. 3 .

## III. EVIDENCE BY SIMULATION

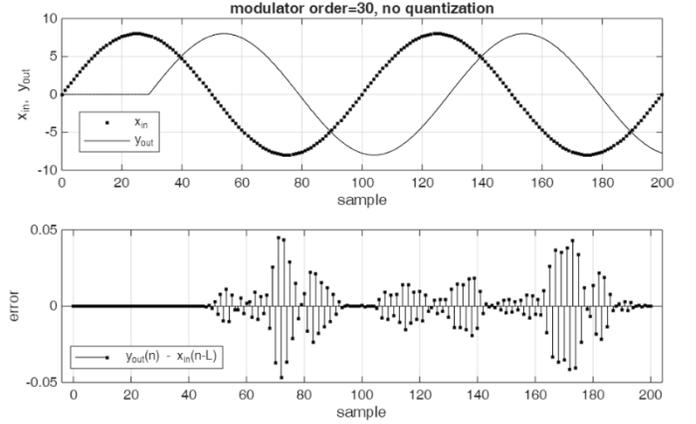

Fig. 8.  Matlab simulation of ΔΣ modulator acc. to Fig. 4, order $L$=30, no quantization ($\Delta_q = 0$), all data represented by floating point numbers.

In the following, the Bounded-Input Bounded Output (BIBO) stability criterion is used when the modulator output is not limited. Otherwise, instability is shown by long series of maximum or minimum output values. The system is BIBO stable if a constant $K$ exists, such that for an input signal $x$ with $|x| < B$ the following applies to the output signal: $|y| < K \cdot B$.

Fig. 8 illustrates a *Matlab* [7] model of the structure proposed in Fig. 4 with order $L = 30$ and consequently 31 coefficients. 201 time points were simulated. All signals and coefficients are represented as floating-point numbers. Stimulus is $x(n) = A \cdot \sin(2\pi \cdot F \cdot n)$ with $A$=8, $F$=1/100, no quantization ($\Delta_q = 0$). Fig. 8 illustrates that the system is BIBO stable.

Fig. 9 demonstrates the sensitivity to coefficient inaccuracies at order $L = 30$. It is nearly the same simulation as in Fig. 8, but the maximum coefficient, $c_{16}$, was multiplied by factor $(1+10^{-12})$, corresponding to (155 117 520 + 0.000155). Although there is no quantization, the stability of the system is lost. As accuracies of $10^{-12}$ are hardly achievable with analog circuitry, this order is most probably applicable for ΔΣ-DDCs only.

Fig. 10 shows the simulation in Fig. 8 with 2 differences: (i) the order was increased from 30 to 51 and (ii) the input signal is rounded according to $x(n) = \text{round}(A \cdot \sin(2\pi \cdot F \cdot n))$ with $A$=8, $F$=1/100. With integer numbers as input, the noise is zero. The maximum coefficients are $c_{26} = c_{27} = 2.48 \cdot 10^{14}$. This underlines the stability of the proposed coefficient theorems.

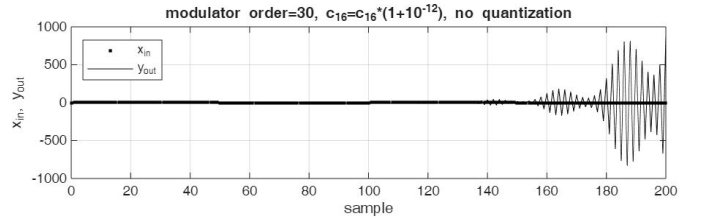

Fig. 9.  Same simulation as Fig. 8, except coefficient c16 = c16·(1+10$^{-12}$)

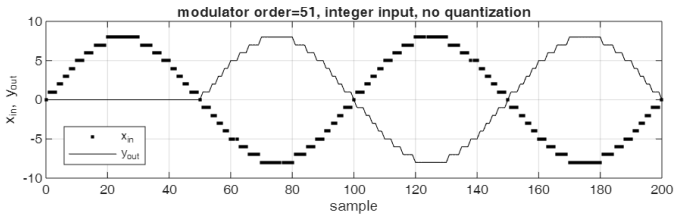

Fig. 10. Same simulation as Fig. 8, except $L$=51 and the input samples are integer numbers. The error vector is not shown because it is all zero.

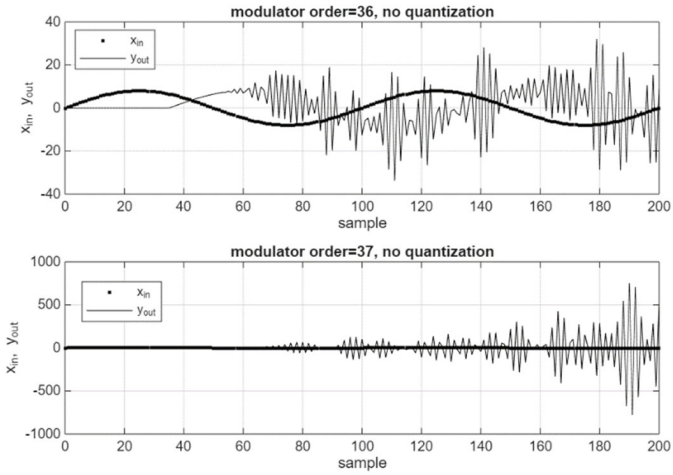

Fig. 11. Simulation like in Fig. 8, except modulator order was increased to 36 (upper part) and 37 (lower part).

Fig. 11 shows a similar simulation as Fig. 8, except with its order increased to $L$=36 in the upper and $L$=37 in the lower part of the picture. For order $L$=36 the system is still stable, for $L$=37 stability is lost. The maximum coefficients for $L$=37 are $c_{19} = c_{20} = 1.767 \cdot 10^{10}$.

Fig. 12 illustrates the results of the simulation of an $L$=10$^{th}$ order ΔΣ-digital-to-digital converter (DDC) intended to lower the resolution of the input data stream, for example for a subsequent DAC as illustrated in Fig. 1(d), a data channel with lower resolution but speed reserves, or for creating "blue" noise. With its input data range of $\pm 2^{16}$ the input width is 17 bit. It is reduced by 8 bits through the quantizer but needs 1 more bit for quantization noise, so that the effective reduction is 7 bits at an oversampling ratio of $OSR$ = 4 and 5, as illustrated in Fig. 12.

TABLE I. DATA USED FOR FIG. 12

| Symbol | Description with respect to Fig. 12 | |
|---|---|---|
| | description | content / details |
| $L$ | Modulator order | 10 |
| $A$ | input signal amplitude | 2^16 = 65 536 |
| $F$ | Relative frequency | = $f/f_s$ = 1/8 and 1/10 |
| $OSR$ | Over-Sampling Ratio | 2/$F$ = 4 and 5 |
| $NoS$ | Number of Samples | 2^24 = 16 777 216 |
| $win$ | window function on $y$ | Blackman-Harris |
| $x(n)$ | input signal | =$A(\sin(\Omega_1 \cdot n)+\sin(\Omega_2 \cdot n))$ |
| $y(n)$ | output signal | |
| $\Delta_q = d_q$ | quantization step | 2^8 = 256 |
| $e_q(n)$ | quantization error | =y(n) – w(n) acc. to Fig. 4 |

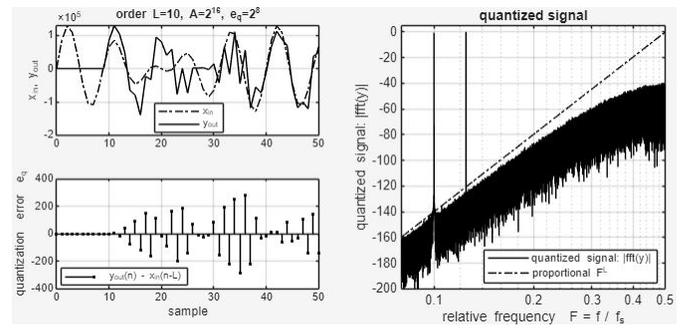

Fig. 12. ΔΣ modulator reducing the bitwidth of a 17-bit data stream by 7 bits. Upper left:16 bit wide input (dashed) vs. modulated data (solid). Lower left: quantization error $e_q(n)$, right: quantization noise $Eq$ over frequency; the dashed line indicates the $F^{10}$ slope of the 10$^{th}$ order noise shaper.

SUMMARY AND OUTLOOK

A method using binomial numbers as integer coefficients for a simple ΔΣ modulator structure consisting of cascaded integrators with distributed feedback and distributed feed-in is presented. Modulator orders as high as 51 are achieved without quantization, a more practical modulator with order 10 is presented with quantization. It is demonstrated that such high order modulators require highly accurate coefficients, which can be achieved with integer coefficients. The high intolerance of inaccurate coefficients at high orders makes this modulator particularly suitable for digital-to-digital converters (DDCs) operating with integer numbers.

In the studies conducted, the resolution of the data stream from input to output could be reduced from 17 to 10 bits using a 10$^{th}$-order modulator at an oversampling ratio of 4. In its current form, the modulator is intolerant of output overloading. Future studies might investigate improved feed-forward input or use resonators as described in [2] to allow for output overloading.

ACKNOWLEDGMENT

The author thanks … for proofreading and helpful suggestions and discussions.